\documentclass[conference,a4paper]{IEEEtran}
\usepackage[latin1]{inputenc}
\usepackage{graphicx,subfigure}
\usepackage{amsmath,amssymb,latexsym,pslatex,dsfont}
\usepackage{epsfig}
\usepackage{multirow}
\usepackage{xspace}
\usepackage{color,colortbl}
\usepackage{array}
\usepackage{cite}
\usepackage{balance}

\IEEEoverridecommandlockouts

\newcommand{\cf}{\mbox{\textit{cf.}}\xspace}
\newcommand{\eg}{\mbox{\textit{e.g.,}}\xspace}
\newcommand{\ie}{\mbox{\textit{i.e.,}}\xspace}
\newcommand{\al}{\mbox{\textit{et al.}}\xspace}
\newcommand{\qf}{$\mathcal{Q}$-factor\xspace}

\newenvironment{slist}%
{ \begin{list}%
	{$\bullet$}%
	{\setlength{\labelwidth}{-5pt}%
	 \setlength{\leftmargin}{0pt}%
	 \setlength{\itemsep}{0pt}}}%
{ \end{list} }
\newcommand{\RANGE}[3]{\mbox{$#1=#2\cdots#3$}\xspace}
\newcommand{\range}[3]{{\RANGE{#1}{#2}{#3}}}
\newcommand{\ALLRANGE}[3]{\mbox{$\forall#1=#2\cdots#3$}\xspace}
\newcommand{\allrange}[3]{{\ALLRANGE{#1}{#2}{#3}}}
\newcommand{\node}[1]{\mbox{$\mathfrak{u}_{#1}$}\xspace}
\newcommand{\link}[1]{\mbox{$\mathfrak{e}_{#1}$}\xspace}
\newcommand{\src}[1]{\mbox{$\mathfrak{s}_{#1}$}\xspace}
\newcommand{\dst}[1]{\mbox{$\mathfrak{r}_{#1}$}\xspace}

\DeclareGraphicsExtensions{{.png},{.pdf},{.eps}}
\graphicspath{{./Figures/}}
\usepackage{draftwatermark}
\SetWatermarkText{Under review}
\SetWatermarkScale{0.7}
\SetWatermarkColor[gray]{0.8}
\SetWatermarkAngle{45}

\title{\fontsize{23}{28}\selectfont Exact Approach for Survivable Regenerator Placement in Translucent WDM Networks\thanks{IEEE Copyright notice.}\thanks{A version of this article is under review for a peer-reviewed IEEE journal.}}
\author{\IEEEauthorblockN{Elias A. Doumith}
\IEEEauthorblockA{TICKET Lab - Antonine University\\
B.P. 40016 Hadat-Baabda - Lebanon\\
Email: elias.doumith@ua.edu.lb}
\and
\IEEEauthorblockN{Sawsan Al Zahr}
\IEEEauthorblockA{LTCI CNRS - Telecom ParisTech - Université Paris-Saclay\\
46, rue Barrault F 75634 Paris Cedex 13 - France\\
Email: sawsan.alzahr@telecom-paristech.fr}}
\begin{document}
\maketitle
\begin{abstract}
Most studies addressing translucent network design targeted a tradeoff between minimizing the number of deployed regenerators and minimizing the number of regeneration sites. The latter highly depends on the carrier's strategy and is motivated by various considerations such as power consumption, maintenance and supervision costs. However, concentrating regenerators into a small number of nodes exposes the network to a high risk of data losses in the eventual case of regenerator pool failure. In this paper, we address the problem of survivable translucent network design taking into account the simultaneous effect of four transmission impairments. We propose an exact approach based on a mathematical formulation to solve the problem of regenerator placement while ensuring the network survivability in the hazardous event of a regenerator pool failure. For this purpose, for each accepted request requiring regeneration, the network management plane computes in advance several routing paths along with associated valid wavelengths going through different regeneration sites. In doing so, we target to implement an $M:N$ shared regenerator pool protection scheme. Simulation results highlight the gain obtained by reducing the number of regeneration sites without sacrificing network survivability.
\end{abstract}

\section{Introduction}%
Over the last two decades, a great attention has been paid to physical layer impairments occurring in long-haul WDM (Wavelength Division Multiplexing) networks. Many research efforts focused on the analysis of the quality of the optical signal in WDM transmission systems operating at high bit-rates (\eg 10, 40, 100 Gbps). Indeed, experiment results pinpointed that according to the current state of technology, transmission impairments induced by long-haul optical equipment may significantly degrade the quality of the optical signal \cite{SCHMIDT:2008}. We distinguish between linear and nonlinear impairments. Linear impairments are proportional to the traveled distance and depend on the signal itself (\eg chromatic dispersion, amplified spontaneous emission noise, polarization mode dispersion), while nonlinear impairments arise from the interaction between neighboring channels (\eg self-phase modulation, cross-phase modulation, four-wave mixing).

In order to cope with the physical impairments and to extend the signal reach, 3R (Re-amplification, Re-shaping and Re-timing) regeneration must be performed at some intermediate nodes along the optical path so that the signal quality is sufficient at its destination. In this respect, translucent WDM networks stand mid-way between opaque WDM networks where 3R regeneration is performed systematically at each switching node, and all-optical WDM networks where the optical signal remains in the optical domain without undergoing any electronic processing at intermediate nodes. Previous studies showed that deploying regenerators into a limited number of nodes (\ie translucent networks) can achieve admissible quality of transmission (QoT) comparable to those obtained in networks with full-regeneration capabilities (\ie opaque networks) \cite{SHEN:2002,YOUSSEF:2011,YANG:2005, ALZAHR_ConTEL:2007,PACHNICKE:2008,PAN:2008,ZHANG:2009,MANOU-JLT:2009,DOUMITH_ICC:2011,ALZAHR_ICT:2011,GAGNAIRE_ONDM:2011}. A comprehensive survey of studies carried out in this domain is provided in \cite{AZODOLMOLY:2009}.

In this matter, two research trends can be distinguished namely, topology-driven and traffic-driven regenerator placement approaches. On the one hand, in topology-driven approaches \cite{SHEN:2002,YOUSSEF:2011}, a few number of nodes are chosen as regeneration sites.
This is achieved by selecting the nodes with the highest number of shortest-paths traversing them. In contrast with \cite{SHEN:2002}, additional regeneration sites may be deployed during the routing and wavelength assignment phase in order to maximize the number of satisfied requests while minimizing the number of regenerators and regeneration sites \cite{YOUSSEF:2011}.
On the other hand, traffic-driven approaches \cite{YANG:2005,ALZAHR_ConTEL:2007,PACHNICKE:2008,PAN:2008,ZHANG:2009,MANOU-JLT:2009, DOUMITH_ICC:2011,ALZAHR_ICT:2011,GAGNAIRE_ONDM:2011} are based on the knowledge of traffic forecasts. Early studies carried out in the field of translucent network design considered either permanent or semi-permanent lightpath requests. It was until 2011, that the problem of regenerator placement has been first investigated under dynamic but deterministic traffic requests \cite{DOUMITH_ICC:2011}. Under such a traffic pattern, one can take advantage from the dynamics of the traffic model so that deployed regenerators may be shared among multiple time-disjoint requests.

In the early 2000s, all studies that addressed the regenerator placement problem were based on empirical laws or heuristic approaches \cite{YANG:2005,ALZAHR_ConTEL:2007,PACHNICKE:2008}. It was until 2008, that Pan \al proposed in \cite{PAN:2008} the first exact approach for regenerator placement under $1+1$ protection scheme minimizing the number of regeneration sites. In this study, the authors formulate the QoT constraint as a maximum all-optical signal reach. Later on, two exact approaches were proposed taking into account different linear and nonlinear impairments \cite{ZHANG:2009,MANOU-JLT:2009}. In these studies, the problem of regenerator placement was formulated as a virtual topology design problem where the QoT constraint was implemented as a minimum admissible \qf. In \cite{ZHANG:2009}, the network topology is represented by an equivalent graph where two non-adjacent nodes, interconnected by a path with an admissible \qf, are connected by a crossover edge in the equivalent graph. Considering static traffic requests, the objective of the proposed approach is to minimize the number of regeneration sites. In \cite{MANOU-JLT:2009}, a set of pre-computed paths is selected and regenerators are deployed along these paths. By routing a set of static traffic requests on the virtual topology, the aim of the proposed approach is to minimize either the number of regeneration sites or the total number of regenerators.

Minimizing the number of deployed regenerators and regeneration sites is mainly motivated by restrictions on capital and operational expenditures (CapEx/OpEx) \cite{YOUSSEF:2011,PACHNICKE:2008,ZHANG:2009}. However, concentrating regenerators into a small number of nodes may expose the network to a high risk of data losses where some requests would be dropped in the case of a regenerator pool failure. In previous investigations \cite{DOUMITH_ICC:2011,ALZAHR_ICT:2011,GAGNAIRE_ONDM:2011}, we proposed exact approaches for translucent network design under dynamic but deterministic traffic model. Our aim was to maximize the number of established requests while minimizing the number of regenerators and the number of regeneration sites. Based on the carrier's strategy, one can tune the objective function in order to stress the reduction in the number of regenerators and/or the number of regeneration sites.

In this paper, we investigate the survivability of translucent networks in the hazardous event of a regenerator pool failure. For this purpose, for each accepted request and each failure scenario, we compute in advance an alternative routing path along with a valid wavelength using different regeneration sites. Our aim is to achieve an $M:N$ shared regenerator pool protection scheme, thus minimizing the number of regenerators and regeneration sites without sacrificing network survivability.

The remainder of this paper is organized as follows. In Section \ref{Network_Model}, we present a brief description of the investigated scenarios. Our approach of survivable translucent network design is provided in Section \ref{Problem_Formulation} followed in Section \ref{Numerical_Results} by an analysis of the numerical results. Finally, we draw our conclusions in Section \ref{Conclusions}. %
\section{Investigated Scenarios}\label{Network_Model}%

\subsection{Network Environment}%

\begin{figure}
\centering
\includegraphics[width = 0.70\columnwidth]{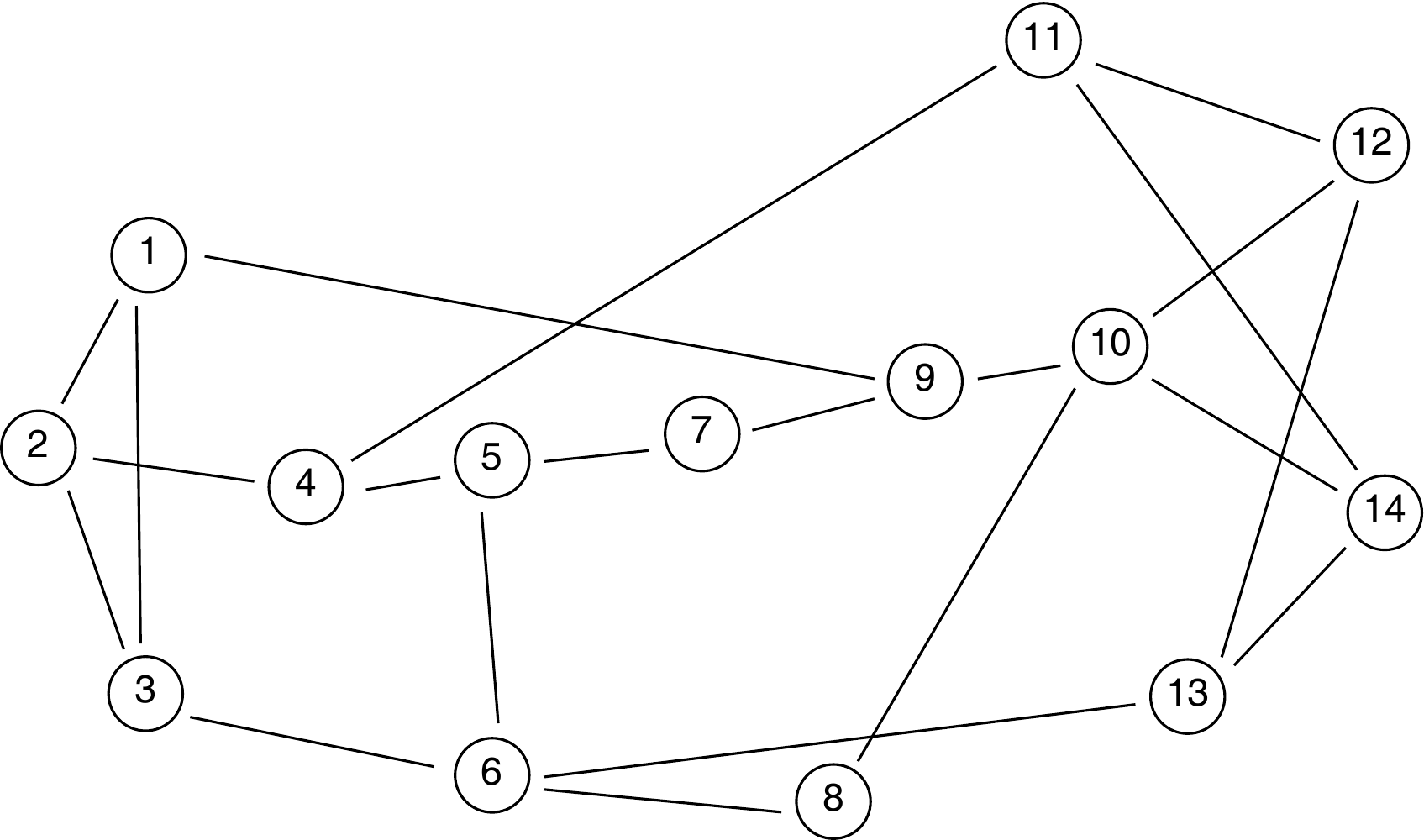}\\\vspace{2pt}
{\tiny\setlength{\extrarowheight}{1.5pt}
\begin{tabular}{@{~}c@{~}|@{~}c@{~}p{10pt}@{~}c@{~}|@{~}c@{~}p{10pt}@{~}c@{~}|@{~}c@{~}p{10pt}@{~}c@{~}|@{~}c@{~}}
Links & Dist. (km) & & Links & Dist. (km) & & Links & Dist. (km) & & Links & Dist. (km) \\
\cline{1-2}\cline{4-5}\cline{7-8}\cline{10-11}
\node{1}-\node{2}  & 480  &&
\node{1}-\node{3}  & 680  &&
\node{1}-\node{9}  & 1500 &&
\node{2}-\node{3}  & 480  \\
\node{2}-\node{4}  & 680  &&
\node{3}-\node{6}  & 850  &&
\node{4}-\node{5}  & 300  &&
\node{4}-\node{11} & 1500 \\
\node{5}-\node{6}  & 480  &&
\node{5}-\node{7}  & 400  &&
\node{6}-\node{8}  & 850  &&
\node{6}-\node{13} & 1500 \\
\node{7}-\node{9}  & 400  &&
\node{8}-\node{10} & 620  &&
\node{9}-\node{10} & 400  &&
\node{10}-\node{12} & 480  \\
\node{10}-\node{14} & 680  &&
\node{11}-\node{14} & 400  &&
\node{12}-\node{13} & 680  &&
\node{13}-\node{14} & 400  \\
\cline{1-2}\cline{4-5}\cline{7-8}\cline{10-11}
\end{tabular}}
\caption{The north American 14-node 20-link NSF backbone network.} \label{Fig:NetworkTopology}\vspace{-10pt}
\end{figure}

For our evaluation, we have considered the $14$-node $20$-link NSF network (Figure \ref{Fig:NetworkTopology}). A network node is a wavelength selective switch-based optical cross connect (WSS-OXC) that can be equipped with a pool of regenerators. These regenerators are responsible for re-amplifying, re-shaping and re-timing the optical signal as well as performing wavelength conversion. A network link is composed by two unidirectional standard single mode fibers (one SMF in each direction) carrying each $W=20$ wavelengths in the C-band. In order to compensate the attenuation caused by fiber losses and the chromatic dispersion, double stage Erbium-doped fiber amplifiers (EDFAs) are deployed every $80$ km along with dispersion compensating fibers (DCFs). Furthermore, inline optical gain equalizers are deployed every $400$ km. Table \ref{Table:param} summarizes the parameters of all the equipment deployed in the network.

A prediction tool, referred to as BER-Predictor, is used to estimate the signal quality at the end of a lightpath \cite{ALZAHR_ConTEL:2007}. BER-Predictor computes the \qf as a function of the penalties simultaneously induced by four physical impairments, namely amplified spontaneous emission noise, chromatic dispersion, polarization mode dispersion and self-phase modulation. The analytical relation between the \qf and the aforementioned impairments has been derived from both analytical formulas and experimental measurements \cite{MOREA:2008}.

%
%

\begin{table*}
\caption{Transmission system parameters} \label{Table:param}
\centering
\setlength{\extrarowheight}{3pt}
\begin{tabular}{@{}l|l@{}p{12mm}@{}l|l@{}p{12mm}@{}l|l@{}}
\cline{1-2} \cline{4-5} \cline{7-8}
Parameter & Value & & Parameter & Value & & Parameter & Value\\
\cline{1-2} \cline{4-5} \cline{7-8}
\multicolumn{8}{l}{} \vspace{-10pt}\\
\cline{1-2} \cline{4-5} \cline{7-8}
Number of wavelengths & $20$ & & SMF PMD (ps/$\sqrt{km}$) & $0.1$ & & Switching losses (dB) & $-13$ \\
\cline{1-2} \cline{4-5} \cline{7-8}
Wavelengths (nm) & $1538.97-1554.13$ & & SMF dispersion (ps/nm.km) & $\overset{av.}{=}17^1$ & &  Inline EDFA Noise Figure (dB) & $\overset{av.}{=}6^1$\\
\cline{1-2} \cline{4-5} \cline{7-8}
Channel spacing (GHz) & $100$ & & DCF input power (dBm) & $-7$ & & Booster EDFA Noise Figure (dB) & $\overset{av.}{=}5.25^1$\\
\cline{1-2} \cline{4-5} \cline{7-8}
Channel bit rate (Gbps) & $10$ & & DCF losses (dB/km) & $0.6$ & & Pre-compensation $(ps)$ & $-800$ \\
\cline{1-2} \cline{4-5} \cline{7-8}
SMF input power (dBm) & $-1$ & & DCF dispersion (ps/nm.km) & $\overset{av.}{=}-90^1$ & & Dispersion slope $(ps/nm/span)$ & $100$ \\
\cline{1-2} \cline{4-5} \cline{7-8}
SMF losses (dB/km) & $0.23$ & & DCF PMD (ps/$\sqrt{km}$) & $0.08$ & & \qf threshold $(dB)$ & $15.6$ \\
\cline{1-2} \cline{4-5} \cline{7-8}
\multicolumn{8}{l}{} \vspace{-11pt}\\
\multicolumn{8}{l}{\scriptsize{${}^1$ ~It is only the mean value; the real value depends on the selected wavelength value. \vspace{-6pt}}}
\end{tabular}
\end{table*}

\subsection{Traffic Model}%

Our proposed model needs to be evaluated under long-term traffic requests as well as dynamic requests. For this purpose, we considered the well-known scheduled lightpath demands (SLD) model where the $i^{th}$ request $\delta_i$ is represented by the tuple $(\src{i},\dst{i},\alpha_{i},\beta_{i},\kappa_{i})$. The source \src{i} and destination \dst{i} nodes of a request are chosen uniformly among the network nodes such that there is no demand between adjacent nodes. The idea is to exclude one-hop lightpaths that do not require any regeneration. The parameters $\alpha_{i}$ and $\beta_{i}$ denote the set-up and tear-down dates of a request, while the parameter $\kappa_{i}$ corresponds to its requested traffic rate. Without loss of generality, we assume that each SLD require the capacity of an optical channel thus, $\kappa_i=1$.

In our simulation, we first assume that all the requests arrive at the same time ($\alpha_i = 0,~\forall i$) and, if accepted, will hold the network for the whole simulation period ($\beta_i = \Delta,~\forall i$). Such requests are known as permanent lightpath demands (PLDs). Without changing the source and destination nodes of the requests, we then reduce the period where they are active according to a parameter $\pi$ ($0 \leqslant\pi\leqslant 1$). More precisely, the activity period $(\beta_{i}-\alpha_{i})$ of a request $\delta_i$ is chosen uniformly in the interval $[\Delta\times\pi-1, \Delta\times\pi+1]$, and the set-up date $\alpha_{i}$ is chosen randomly while ensuring that $\delta_i$ still ends before the expiration of the simulation period ($\beta_i \leqslant \Delta$).
\section{Mathematical Formulation for Survivable Impairment-Aware Network Planning}\label{Problem_Formulation}%

In this paper, we propose an exact approach to solve the regenerator placement problem while ensuring the survivability of the network in the hazardous event of a regenerator pool failure. This can be achieved by formulating the problem as a Mixed-Integer Linear Program (MILP) and solving it using traditional solvers. To that end, we consider different sets of dynamic but predictable traffic requests. Thanks to their predictability, these traffic requests can be routed off-line by the management plane. By examining the network topology and the traffic requests, the management plane has to decide which requests will be accepted and which ones will be rejected. For each accepted request, the management plane will select a suitable routing path along with a valid wavelength and a set of regenerators. These regenerators are required in order to cope with transmission impairments and for wavelength conversion needs. Furthermore, for each eventual failure of a regenerator pool, the management plane must compute in advance, for each accepted request, an alternative routing path along with a new valid wavelength and a new set of regenerators.

In order to improve the scalability of our approach, we decompose the problem into the ``Routing and Regenerator Placement'' (RRP) sub-problem and the ``Wavelength Assignment and Regenerator Placement'' (WARP) sub-problem. The common parameters for these two sub-problems are:
\begin{slist}
\item The network topology represented by a graph $\mathcal{G}=(\mathcal{V},\mathcal{E})$, where $\mathcal{V} = \{ \node{v}, \range{v}{1}{N}\}$ is the set of nodes, and $\mathcal{E} = \{\link{e}=(\node{v},\node{u}) \in \mathcal{V}\times\mathcal{V}, \range{e}{1}{L}\}$ is the set of unidirectional fiber-links connecting the nodes.
\item The set of wavelengths $\Lambda =\{\lambda_{\ell}, \range{\ell}{1}{W}\}$ available on each fiber-link in the network.
\item  The threshold $\mathcal{Q}_{th}$ for an admissible \qf.
\end{slist}

As we are concerned by the failure of a regenerator pool that could be located at any node of the network, we consider for each of the sub-problems $N+1$ different scenarios. Scenario `0' corresponds to the case where all the regenerator pools are fully operational, while scenario `$s$' (\RANGE{s}{1}{N}) corresponds to the case where the regenerator pool at node $\node{s}$ is down.


\subsection{Routing and Regenerator Placement}%

In this sub-problem, we aim at maximizing the number of satisfied requests while minimizing the number of regeneration sites as well as the number of regenerators. For this purpose, we assume that the quality of transmission is independent of the wavelength value. In other words, the QoT of a lightpath transmitted over a wavelength $\lambda$ is the same as if the lightpath was transmitted over the reference wavelength $\lambda_{c}=1550$ nm. The RRP sub-problem is formulated as follows.

\subsubsection{Parameters}%

\begin{slist}
\item The set of requests $\mathcal{D}=\{\delta_{i}, \range{i}{1}{D}\}$. Each request $\delta_{i}$ is represented by a tuple $(\src{i}\in \mathcal{V},\dst{i}\in \mathcal{V},\alpha_{i},\beta_{i},\kappa_{i}=1)$.
\item For each request, we compute beforehand $K$-alternate shortest paths in terms of effective length connecting its source node to its destination node. Let $\mathcal{P}_{i}=\{p_{i,j}, \range{j}{1}{K}\}$ be the set of available routes for the request $\delta_{i}$. The $j^{th}$-shortest path $p_{i,j}$ of $\delta_{i}$ is the ordered set of unidirectional links $\{\link{e_1}, \link{e_2}, \cdots, \link{e_{|p_{i,j}|}}\}$ traversed in the source-destination direction ($\src{i}\mapsto\dst{i}$). For each link pair $(\link{m}, \link{n})$ along a path $p_{i,j}$, we compute by means of BER-Predictor the \qf value $\mathfrak{Q}_{i,j}^{m,n}$ of the directed path-segment delimited by the \emph{source} node of link $\link{m}$ and the \emph{destination} node of link $\link{n}$ ($\link{m} \leqslant \link{n}$).
\item The ordered set $\mathcal{T}$ grouping the set-up and tear-down times of all the requests in $\mathcal{D}$.
\begin{eqnarray}
\mathcal{T} = \bigcup\limits_{\delta_i\in\mathcal{D}} \{\alpha_{i}, \beta_{i}\} = \{ \tau_{1}, \cdots, \tau_{\mathfrak{T}}\}&&\\ \nonumber \textrm{such that } \tau_{1} < \tau_{2} < \cdots < \tau_{\mathfrak{T}} \textrm{ and } \mathfrak{T}=|\mathcal{T}|\quad\quad&&\qquad\qquad
\end{eqnarray}
\item The request matrix $\Theta=\{\theta_{i,t}, \range{i}{1}{D}, \range{t}{1}{\mathfrak{T}}\}$ representing the traffic requests over time. An element $\theta_{i,t}$ of this matrix is a binary value specifying the presence $(\theta_{i,t}=1)$ or the absence $(\theta_{i,t}=0)$ of request $\delta_{i}$ at time instant $\tau_{t}$.
\begin{equation}
   \theta_{i,t}=
      \begin{cases}
      1 & \textrm{if $\alpha_i\leqslant\tau_{t}<\beta_i$,}\\
      0 & \textrm{otherwise.}%
   \end{cases}\label{Eq:RequestMat}
\end{equation}
\end{slist}

\subsubsection{Variables}%

\begin{slist}
\item The binary variables $\mathfrak{a}_{i}$, \range{i}{1}{D}.\\
    $\mathfrak{a}_{i}=1$, if the traffic request $\delta_i$ is accepted. $\mathfrak{a}_{i}=0$, otherwise.
\item The binary variables $\mathfrak{p}_{s,i,j}$, \range{s}{0}{N}, \range{i}{1}{D}, \range{j}{1}{K}.\\
    $\mathfrak{p}_{s,i,j}=1$, if in the selected scenario `$s$', the $j^{th}$-shortest path between \src{i} and \dst{i} is assigned to request $\delta_i$. $\mathfrak{p}_{s,i,j}=0$, otherwise.
\item The binary variables $\zeta_{s,i,j}^{m,n}$, \range{s}{0}{N}, \range{i}{1}{D}, \range{j}{1}{K}, \range{m}{1}{L}, \range{n}{1}{L}.\\
    $\zeta_{s,i,j}^{m,n}$ is an intermediate variable used to insure that the value of the \qf at the end of the path-segment delimited by the source node of link \link{m} and the destination node of link \link{n} along the path $p_{i,j}$ used by the request $\delta_i$ in the scenario `$s$' exceeds the predefined threshold.
\item The binary variables $\mathfrak{d}_{s,i,u}$, \range{s}{0}{N}, \range{i}{1}{D}, \range{u}{1}{N}.\\
    $\mathfrak{d}_{s,i,u}=1$, if in the scenario `$s$', the request $\delta_i$ is regenerated at node \node{u}. $\mathfrak{d}_{s,i,u}=0$, otherwise.
\item The regenerator matrix $\Psi=\{\psi_{s,u,t}, \range{s}{0}{N}, \range{u}{1}{N},$ $\range{t}{1}{\mathfrak{T}}\}$.\\
    $\psi_{s,u,t}$ is a non-negative integer variable equal to the number of regenerators that are in use in the scenario `$s$' at node \node{u} at time instant $\tau_t$.
\item The binary variables $\phi_{u}$, \range{u}{1}{N}.\\
    $\phi_{u}=1$, if the node \node{u} is a regeneration site. $\phi_{u}=0$, otherwise.
\item The non-negative integer variables $\mathfrak{R}_{u}$, \range{u}{1}{N}.\\
    $\mathfrak{R}_{u}$ denotes the number of regenerators deployed at node \node{u}.
\end{slist}

\subsubsection{Constraints}%

\begin{slist}
\item If the request $\delta_i$ is accepted, it must be routed over a unique path among the available $K$-shortest paths in each of the considered scenario `$s$'. \allrange{s}{0}{N},~\allrange{i}{1}{D},
\begin{equation}
   \sum_{j=1\cdots K}\mathfrak{p}_{s,i,j} = \mathfrak{a}_{i}\label{eq::poorPerformance}
\end{equation}
\item In each scenario `$s$', the number of requests routed over a single fiber-link must not exceed the number of wavelengths on that link. \allrange{s}{0}{N},~\allrange{t}{1}{\mathfrak{T}},~\allrange{m}{1}{L},
\begin{equation}
   \sum_{i=1\cdots D}~\sum_{\begin{subarray}{l}
   j=1\cdots K\\
   \backslash \mathfrak{e}_{m} \in p_{i,j}
   \end{subarray}}\theta_{i,t}\times\mathfrak{p}_{s,i,j}\leqslant W
\end{equation}
\item The value of the \qf at the end of the path-segment delimited by any two distinct nodes along the selected path of an accepted request must exceed the predefined threshold $\mathcal{Q}_{th}$. Otherwise, regenerators must be deployed at some intermediate nodes along this path-segment. \allrange{s}{0}{N},~\allrange{i}{1}{D},~\allrange{j}{1}{K},~$\forall \mathfrak{e}_{n} \in p_{i,j}$,
\begin{eqnarray}
   \sum_{\begin{subarray}{l}
   \mathfrak{e}_{m} \in p_{i,j}\\
   \backslash \mathfrak{e}_{m} \leqslant \mathfrak{e}_{n}
   \end{subarray}}\zeta_{s,i,j}^{m,n}\times\mathfrak{Q}_{i,j}^{m,n} & \geqslant & \mathfrak{p}_{s,i,j}\times\mathcal{Q}_{th}\\
   \sum_{\begin{subarray}{l}
   \mathfrak{e}_{m} \in p_{i,j}\\
   \backslash \mathfrak{e}_{m} \leqslant \mathfrak{e}_{n}
   \end{subarray}}\zeta_{s,i,j}^{m,n} & = & \mathfrak{p}_{s,i,j}
\end{eqnarray}
\item By collecting all the previous constraints on the variables $\zeta_{s,i,j}^{m,n}$, we can determine all the intermediate nodes \node{u} where a request $\delta_i$ should be regenerated (except at its source node \src{i}). \allrange{s}{0}{N},~\allrange{i}{1}{D},~\allrange{j}{1}{K}, $\forall \mathfrak{e}_{m}=(\mathfrak{u}_{u}, \mathfrak{u}_{v})\in p_{i,j}$, $\forall \mathfrak{e}_{n}\in p_{i,j}$ such that $\mathfrak{e}_{m}\leqslant \mathfrak{e}_{n}$ and $\mathfrak{u}_{u} \neq \mathfrak{s}_{i}$,
\begin{equation}
   \mathfrak{d}_{s,i,u} \geqslant \zeta_{s,i,j}^{m,n}
\end{equation}
\item The number of regenerators $\psi_{s,u,t}$ in use at node \node{u} at time instant $\tau_t$ for a given scenario `$s$' can then be computed as: \allrange{s}{0}{N},~\allrange{u}{1}{N},~\allrange{t}{1}{\mathfrak{T}},
\begin{equation}
   \psi_{s,u,t} = \sum_{i=1\cdots D}\theta_{i,t}\times\mathfrak{d}_{s,i,u}
\end{equation}
\item The number of regenerators $\mathfrak{R}_{u}$ deployed at node \node{u} is the maximum number of regenerators that are in use at any time for all the considered scenarios. \allrange{s}{0}{N},~\allrange{u}{1}{N},~\allrange{t}{1}{\mathfrak{T}},
\begin{equation}
   \mathfrak{R}_{u} \geqslant \psi_{s,u,t}
\end{equation}
\item A node is considered as a regeneration site if it hosts at least a single regenerator. \allrange{u}{1}{N},
\begin{equation}
   \phi_{u} \geqslant 10^{-3}\times \mathfrak{R}_{u}
\end{equation}
\item Finally, a regenerator pool failure is obtained by setting to zero the number of regenerators that can be deployed at a given node for the corresponding scenario. \allrange{s}{1}{N},~\allrange{t}{1}{\mathfrak{T}},
\begin{equation}
   \psi_{s,s,t} = 0
\end{equation}
\end{slist}

\subsubsection{Objective}%

The objective of the RRP sub-problem is to maximize the number of accepted requests while minimizing the number of regenerators and regeneration sites. This objective is expressed as:
\begin{equation}
\max~~\gamma_1\times\!\!\!\sum_{i=1\cdots D}\!\!\!\mathfrak{a}_{i} - \gamma_2\times\!\!\!\sum_{u=1\cdots N}\!\!\!\phi_{u} - \gamma_3\times\!\!\!\sum_{u=1\cdots N}\!\!\!\mathfrak{R}_{u}
\end{equation}
\noindent where $\gamma_1$, $\gamma_2$, and $\gamma_3$ are three positive real numbers used to stress the regenerators concentration into a limited number of regeneration sites, the minimization of the required number of regenerators, the maximization of the number of accepted requests, or any combination of the previous objectives.

\subsubsection{Performance Improvement}
Although the previous formulation is correct, the feasible solution space is quite large. In order to shorten the time to solve this MILP formulation, we decided to reduce the solution space while paying attention to not omit the optimal solution. This can be achieved by cutting regions of the solution space that do not contain any improvement. Indeed, if we notice that when a node \node{s} is not selected as a regeneration site, the scenario `$s$' representing the failure of the regenerator pool at this node is obvious as it should not affect the accepted requests nor their associated paths. More precisely, the paths assigned to the requests in scenario `$s$' should be identical to the paths obtained in scenario `0'. This can be easily obtained by replacing the previous Equation \eqref{eq::poorPerformance} with the following:
\begin{slist}
\item If the request $\delta_i$ is accepted in the scenario `0', it must be routed over a unique path among the available $K$-shortest paths. \allrange{i}{1}{D},
\begin{equation}
   \sum_{j=1\cdots K}\mathfrak{p}_{0,i,j} = \mathfrak{a}_{i}
\end{equation}
\item For the scenario `$s$', if the node \node{s} is a regeneration site, we have to select a path $p_{i,j}$ for each accepted request $\delta_{i}$. Conversely, if the node \node{s} is not a regeneration site, we will set all the variables $\mathfrak{p}_{s,i,j}$ to zero. In this way, we will not assign any path to the accepted requests. Once we obtain the optimal solution, we will route, in a post-processing step, each accepted request in the scenario `$s$' on the same path as in the scenario `0'. This can be formulated mathematically as: \allrange{s}{1}{N},~\allrange{i}{1}{D},
\begin{equation}
   \sum_{j=1\cdots K}\mathfrak{p}_{s,i,j} = \mathfrak{a}_{i} \times \phi_{s}\label{eq::NonLinear}
\end{equation}
The term $\mathfrak{a}_{i} \times \phi_{s}$ is non-linear since it is the product of the two variables. However, this product can be linearized by means of additional constraints. Thus, Equation \eqref{eq::NonLinear} can be written in linear form as follows: \allrange{s}{1}{N},~\allrange{i}{1}{D},

~\vspace*{-15pt}
\begin{subequations}
  \begin{align}
    \sum_{j=1\cdots K}\mathfrak{p}_{s,i,j} &\leq \mathfrak{a}_{i}\\
    \sum_{j=1\cdots K}\mathfrak{p}_{s,i,j} &\leq \phi_{s}\\
    \sum_{j=1\cdots K}\mathfrak{p}_{s,i,j} &\geq \mathfrak{a}_{i} + \phi_{s} - 1
  \end{align}
\end{subequations}
\end{slist}

\subsection{Wavelength Assignment and Regenerator Placement}%

The requests, that are rejected in the RRP sub-problem, are definitely dropped and removed from the problem. Let $\widehat{\mathcal{D}}=\{\widehat{\delta}_{i}, \range{i}{1}{\widehat{D}}\}$ be the set of the remaining requests that are accepted. Each accepted request has been assigned a unique path for each considered scenario and eventually required to be regenerated at some intermediate nodes along its path. Without altering its selected path, an accepted request requiring regeneration is divided into path-segments for each considered scenario whenever it passes through its regeneration site. As the routes and the positions of the regenerators assigned to a given request may vary from one scenario to another, its decomposition into sub-paths will also vary. Let $\widetilde{\mathcal{D}}_{s}=\{\widetilde{\delta}_{s,d}, \range{d}{1}{\widetilde{D}_{s}}\}$ be the modified sets of requests (one modified set of requests for each considered scenario \range{s}{0}{N}) containing the accepted requests with an admissible QoT (no regeneration required) as well as the path-segments of the accepted requests requiring regeneration.

In this sub-problem, we assign to each request $\widetilde{\delta}_{s,d}$ in a given scenario `$s$' a single continuous wavelength between its source and its destination nodes. When this is not possible, additional regenerators are deployed to serve as wavelength converters. Moreover, all these requests have an acceptable QoT if they are transmitted over the reference wavelength $\lambda_{c}=1550$ nm. If a request $\widetilde{\delta}_{s,d}$ is transmitted over another wavelength, its QoT may be degraded due to the non-flat spectral response of optical components. This problem can be resolved by adding regenerators along the path assigned to $\widetilde{\delta}_{s,d}$. However, it may happen that the required additional regenerator for a given scenario `$s$' needs to be deployed at node \node{s}. Recalling that scenario `$s$' corresponds to the case where the regenerator pool at node $\node{s}$ is down, no regenerators can be deployed at node \node{s} and the corresponding request will be rejected. In order to optimize the utilization of the network ressources, whenever a request $\widetilde{\delta}_{s,d}$ is rejected, we will also reject the initial request $\widehat{\delta}_{i}$ and all its path-segments from all the scenarios. Furthermore, we will also remove all the regenerators that were introduced by the initial request $\widehat{\delta}_{i}$ in the RRP sub-problem. For this purpose, we define the function $\mathfrak{F}(.)$ that for each $\widetilde{\delta}_{s,d} \in \widetilde{\mathcal{D}}_{s}$ returns the index of the associated initial request $\widehat{\delta}_{i} \in \widehat{\mathcal{D}}$.
\begin{equation}
   \mathfrak{F}\left(\widetilde{\delta}_{s,d}\right) = i
\end{equation}

The WARP sub-problem can be formulated as follows.

\subsubsection{Parameters}%

\begin{slist}
\item The set of initial requests $\widehat{\mathcal{D}}=\{\widehat{\delta}_{i}, \range{i}{1}{\widehat{D}}\}$ that were accepted in the RRP sub-problem. Each accepted initial request is routed over a single path and may be regenerated at some intermediate nodes along this path. For this purpose, we define the binary parameters $\widehat{\mathfrak{d}}_{s,i,u}$, \range{s}{0}{N}, \range{i}{1}{\widehat{D}}, \range{u}{1}{N}. $\widehat{\mathfrak{d}}_{s,i,u} = 1$, if the request $\widehat{\delta}_{i}$ was regenerated in the scenario `$s$' of the RRP problem at node \node{u}. $\widehat{\mathfrak{d}}_{s,i,u} = 0$, otherwise.
\item The new sets of requests $\widetilde{\mathcal{D}}_{s}=\{\widetilde{\delta}_{s,d}, \range{d}{1}{\widetilde{D}_{s}}\}$ (\range{s}{0}{N}) obtained by dividing the initial requests at the nodes where they were regenerated. As in the previous sub-problem, each request $\widetilde{\delta}_{s,d}$ is represented by a tuple $(\src{s,d}\in \mathcal{V},\dst{s,d}\in \mathcal{V},\alpha_{s,d},\beta_{s,d},\kappa_{s,d}=1)$.
\item At the end of the RRP sub-problem, each request $\widetilde{\delta}_{s,d}$ is routed over a single path $p_{s,d}$. This path $p_{s,d}$ is defined as the ordered set of unidirectional links $\{\link{e_1}, \link{e_2}, \cdots, \link{e_{|p_{s,d}|}}\}$ traversed in the source-destination direction ($\src{s,d}\mapsto\dst{s,d}$). For each wavelength $\lambda_{\ell}\in\Lambda$, we compute by means of BER-Predictor the \qf value $\mathfrak{Q}_{s,d}^{\ell}$ at the destination node \dst{s,d} of the selected path $p_{s,d}$.
\item The ordered set $\mathcal{T}=\{\tau_{t}, \range{t}{1}{\mathfrak{T}}\}$ grouping the set-up and tear-down times of all the requests $\widetilde{\delta}_{s,d}$ in $\widetilde{\mathcal{D}}_{s}$ (\range{s}{0}{N}).
\item The request matrix $\widehat{\Theta}=\{\widehat{\theta}_{i,t}, \range{i}{1}{\widehat{D}}, \range{t}{1}{\mathfrak{T}}\}$ representing the accepted initial traffic requests $\widehat{\delta}_{i}$ over time.
\item The new request matrices $\widetilde{\Theta}_{s}=\{\widetilde{\theta}_{s,d,t}, \range{d}{1}{\widetilde{D}_{s}},$ $\range{t}{1}{\mathfrak{T}}\}$ (\range{s}{0}{N}) representing for each scenario `$s$' the requests $\widetilde{\delta}_{s,d}$ over time. These matrices are computed in the same way as in the RRP sub-problem (\cf Equation \eqref{Eq:RequestMat}).
\end{slist}

\subsubsection{Variables}%

\begin{slist}
\item The binary variables $\mathfrak{a}_{i}$, \range{i}{1}{\widehat{D}}.\\
    $\mathfrak{a}_{i}=1$, if the initial traffic request $\widehat{\delta}_i$ remains accepted in the WARP sub-problem. $\mathfrak{a}_{i}=0$, otherwise.
\item The binary variables $\varrho_{s,d,m}^{\ell}$, \range{s}{0}{N}, \range{d}{1}{\widetilde{D}_{s}}, \range{m}{1}{L}, \range{\ell}{1}{W}.\\
    $\varrho_{s,d,m}^{\ell}=1$, if in the scenario `$s$', the request $\widetilde{\delta}_{s,d}$ is transmitted over the wavelength $\lambda_{\ell}$ along the link \link{m}. $\varrho_{s,d,m}^{\ell}=0$, otherwise.
\item The binary variables $\mathfrak{d}_{s,d,u}$, \range{s}{0}{N}, \range{d}{1}{\widetilde{D}_{s}}, \range{u}{1}{N}.\\
    $\mathfrak{d}_{s,d,u}=1$, if in the scenario `$s$', the request $\widetilde{\delta}_{s,d}$ is regenerated during the WARP sub-problem at node \node{u}. $\mathfrak{d}_{s,d,u}=0$, otherwise.
\item The new regenerator matrix $\Psi=\{\psi_{s,u,t}, \range{s}{0}{N},$ $\range{u}{1}{N}, \range{t}{1}{\mathfrak{T}}\}$.\\
    $\psi_{s,u,t}$ is a non-negative integer variable equal to the number of regenerators that are in use in the scenario `$s$' at node \node{u} at time instant $\tau_t$.
\item The binary variables $\phi_{u}$, \range{u}{1}{N}.\\
    $\phi_{u}=1$, if the node \node{u} is a regeneration site. $\phi_{u}=0$, otherwise.
\item The non-negative integer variables $\mathfrak{R}_{u}$, \range{u}{1}{N}.\\
    $\mathfrak{R}_{u}$ denotes the total number of regenerators deployed at node \node{u} (including those that were already deployed in the RRP sub-problem).
\end{slist}

\subsubsection{Constraints}%

\begin{slist}
\item If the request $\widehat{\delta}_{i}$ remains accepted, a single wavelength must be reserved on all the links that are traversed by its sub-paths in all the scenarios. \allrange{s}{0}{N},~\allrange{d}{1}{\widetilde{D}_{s}},~\allrange{m}{1}{L},
\begin{equation}
   \sum_{\ell=1\cdots W}\varrho_{s,d,m}^{\ell}=\begin{cases}
      \mathfrak{a}_{\mathfrak{F}\left(\widetilde{\delta}_{s,d}\right)} & \textrm{if } \link{m} \in p_{s,d},\\
      0 & \textrm{otherwise.}
      \end{cases}
\end{equation}
\item For any given scenario `$s$', each wavelength on a link can be used at most once at a given time instant. \allrange{s}{0}{N},~\allrange{m}{1}{L},~\allrange{\ell}{1}{W},~\range{t}{1}{\mathfrak{T}},
\begin{equation}
   \sum_{d=1\cdots \widetilde{D}_{s}}\varrho_{s,d,m}^{\ell} \times \widetilde{\theta}_{s,d,t}\leqslant 1
\end{equation}
\item A path $p_{s,d}$ must use the same wavelength on any two consecutive links unless a regenerator is deployed at the node in common to the two links. \allrange{s}{0}{N},~\allrange{d}{1}{\widetilde{D}_{s}},~\allrange{\ell}{1}{W},~$\forall \link{m}=(\node{v},\node{u}) \in p_{s,d}$,~$\forall \link{n}=(\node{u},\node{l}) \in p_{s,d}$,
\begin{subequations}
   \begin{eqnarray}
   \varrho_{s,d,m}^{\ell} - \varrho_{s,d,n}^{\ell} & \leqslant & \mathfrak{d}_{s,d,u}\\
   \varrho_{s,d,n}^{\ell} - \varrho_{s,d,m}^{\ell} & \leqslant & \mathfrak{d}_{s,d,u}
   \end{eqnarray}
\end{subequations}
\item The value of the \qf at the destination node of a request must exceed the predefined threshold $\mathcal{Q}_{th}$. Otherwise, a regenerator is deployed at any intermediate node along the path of the degraded request. \allrange{s}{0}{N},~\allrange{d}{1}{\widetilde{D}_{s}},~\allrange{\ell}{1}{W},~$\forall \link{m} \in p_{s,d}$,
\begin{equation}
   \mathfrak{Q}_{s,d}^{\ell} \times \varrho_{s,d,m}^{\ell} + \mathcal{Q}_{th}\times \!\!\!\!\!\!\!\!\!\sum_{\begin{subarray}{l}
   \mathfrak{e}_{n}=(\mathfrak{u}_{u},\mathfrak{u}_{v}) \in p_{s,d}\\
   \backslash \mathfrak{u}_{u} \neq \mathfrak{s}_{s,d}
   \end{subarray}} \!\!\!\!\!\!\!\!\!\mathfrak{d}_{s,d,u} \geqslant \mathcal{Q}_{th}\times \varrho_{s,d,m}^{\ell}
\end{equation}
\item The number of regenerators $\psi_{s,u,t}$ in use at node \node{u} at time instant $\tau_{t}$ for a given scenario `$s$' is equal to the number of regenerators that were deployed in the RRP sub-problem for the initial requests that remained accepted augmented by the number of regenerators that are required to serve as wavelength converters and to cope with the QoT degradation due to the non-flat spectral response of optical components. These constraints allow the WARP sub-problem to reuse, when possible, the regenerators deployed during the RRP sub-problem. $\psi_{s,u,t}$ is given by: \allrange{s}{0}{N},~\allrange{u}{1}{N},~\allrange{t}{1}{\mathfrak{T}},
\begin{equation}
   \psi_{s,u,t} = \!\!\!\sum_{i=1\cdots \widehat{D}}\widehat{\theta}_{i,t}\times\widehat{\mathfrak{d}}_{s,i,u}\times\mathfrak{a}_{i} + \!\!\!\!\!\!\sum_{d=1\cdots \widetilde{D}_{s}}\widetilde{\theta}_{s,d,t}\times\mathfrak{d}_{s,d,u}
\end{equation}
\item The number of regenerators $\mathfrak{R}_{u}$ deployed at node \node{u} is the maximum number of regenerators that are in use at any time for all the considered scenarios. \allrange{s}{0}{N},~\allrange{u}{1}{N},~\allrange{t}{1}{\mathfrak{T}},
\begin{equation}
   \mathfrak{R}_{u} \geqslant \psi_{s,u,t}
\end{equation}
\item A node is considered as a regeneration site if it hosts at least a single regenerator. \allrange{u}{1}{N},
\begin{equation}
   \phi_{u} \geqslant 10^{-3}\times \mathfrak{R}_{u}
\end{equation}
\item Finally, a regenerator pool failure is obtained by setting to zero the number of regenerators that can be deployed at a given node for the corresponding scenario. \allrange{s}{1}{N},~\allrange{t}{1}{\mathfrak{T}},
\begin{equation}
   \psi_{s,s,t} = 0
\end{equation}
\end{slist}

\subsubsection{Objective}%

As it was the case for the previous sub-problem, the objective of the WARP sub-problem is to maximize the number of accepted requests while minimizing the number of regenerators and regeneration sites. This objective is expressed as:
\begin{equation}
\max~~\gamma_4\times\!\!\!\sum_{i=1\cdots \widehat{D}}\!\!\!\mathfrak{a}_{i} - \gamma_5\times\!\!\!\sum_{u=1\cdots N}\!\!\!\phi_{u} - \gamma_6\times\!\!\!\sum_{u=1\cdots N}\!\!\!\mathfrak{R}_{u}
\end{equation}
\noindent where $\gamma_4$, $\gamma_5$, and $\gamma_6$ are three positive real numbers used to stress the regenerators concentration into a limited number of regeneration sites, the minimization of the required number of regenerators, the maximization of the number of accepted requests, or any combination of the previous objectives. %
\section{Numerical Results}\label{Numerical_Results}%

In this paper, we aim to emphasize the cost benefit brought by the $M:N$ shared regenerator pool protection scheme compared to the commonly deployed $1+1$ protection scheme. The latter scheme is derived from our previous work \cite{ALZAHR_ICT:2011,DOUMITH_ICC:2011} by assuming that each regeneration site is equipped with two identical pools of regenerators; one for normal operation and the other for backup operation. The optimal results for the two approaches obtained at the end of the WARP phase are compared in terms of average acceptance ratio $\bar{\mathfrak{a}}$ and its standard deviation $\ddot{\mathfrak{a}}$, average number of regeneration sites $\bar{\phi}$ and its standard deviation $\ddot{\phi}$, as well as average number of regenerators $\bar{\mathfrak{R}}$ and its standard deviation $\ddot{\mathfrak{R}}$.

\begin{figure}
\centering
\includegraphics[width =.99\columnwidth]{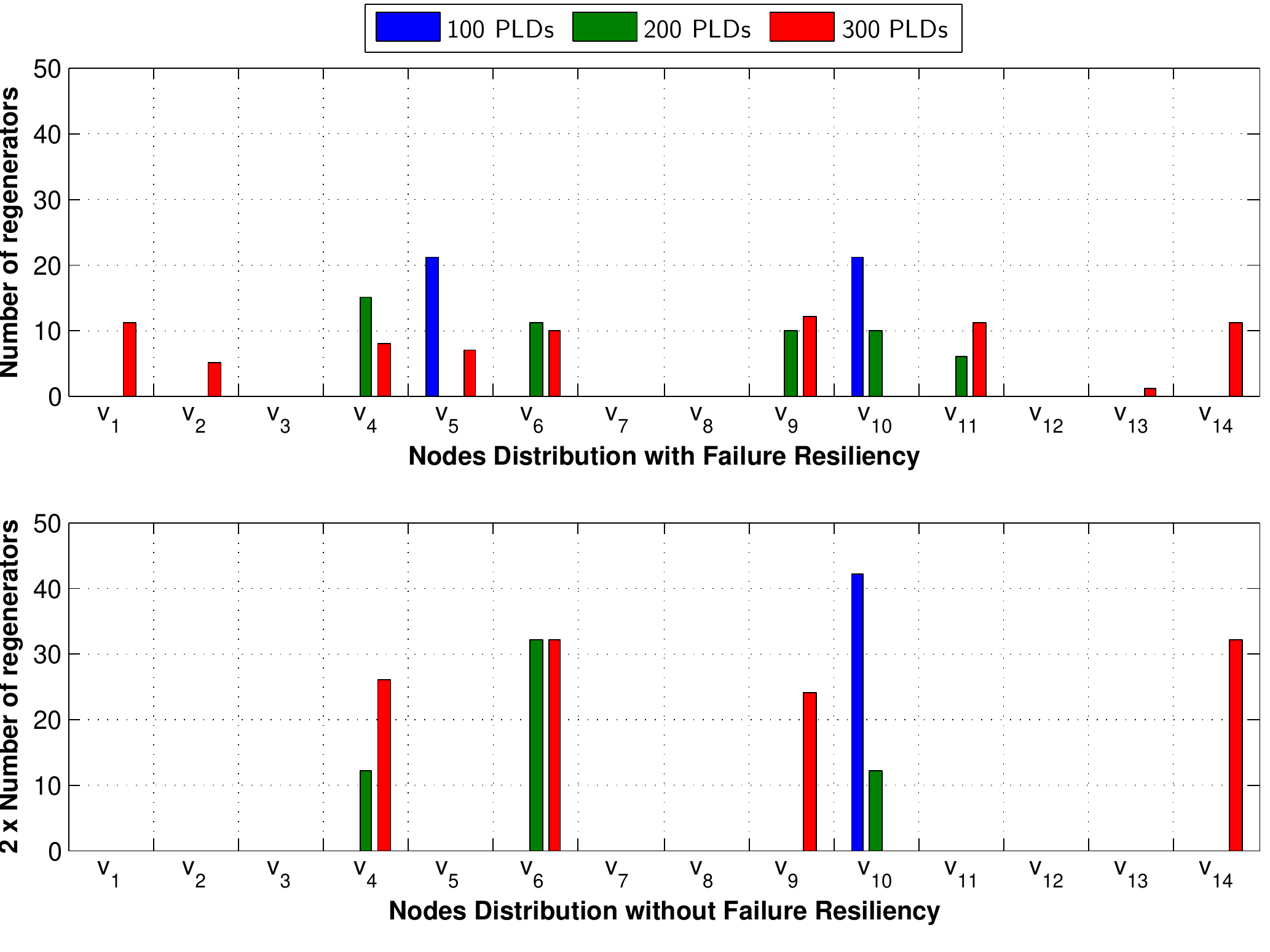}\vspace{-8pt}
\caption{Median regenerator distribution $\mathfrak{R}_{u}$ for various loads of PLDs.} \label{Fig:RegDistributionPLD}
\end{figure}

In a first step, we considered three different loads of permanent requests ($D\in\{100,200,300\}$). For each traffic load, we randomly generated $10$ different sets of PLDs. Table \ref{PLD_Results} summarizes the results obtained for the different traffic loads considered in our evaluation. Figure \ref{Fig:RegDistributionPLD} shows the \emph{median} distribution of the deployed regenerators over the network nodes. It is obvious that the number of regenerators and regeneration sites increase with the traffic load. For $100$ PLDs, the $M:N$ and $1+1$ protection schemes achieve the same results. However, the $M:N$ protection scheme achieves in average a reduction of $22\%$ and $25\%$ in the number of deployed regenerators compared to the $1+1$ protection scheme for the sets of 200 and 300 PLDs, respectively.

\begin{figure}
\centering
\includegraphics[width =.99\columnwidth]{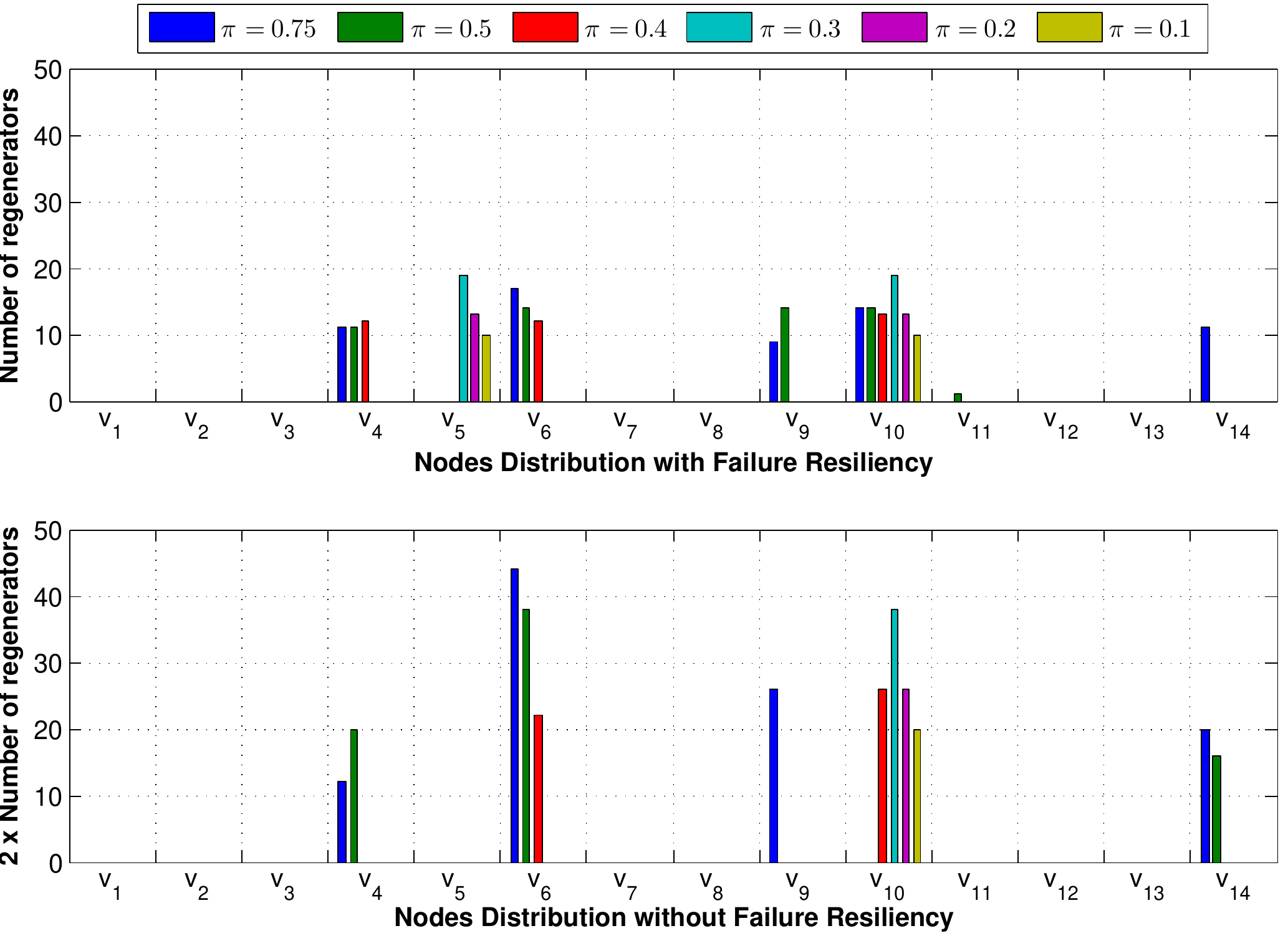}\vspace{-8pt}
\caption{Median regenerator distribution $\mathfrak{R}_{u}$ for various sets of 200 SLDs.} \label{Fig:RegDistributionSLD}
\end{figure}

In a second step, we investigated the impact of the requests' time-correlation on the number of regenerators and regeneration sites by considering dynamic requests with different activity periods ($\pi\in\{0.1,0.2,0.3,0.4,0.5,0.75\}$). For each value of the time-correlation, we randomly generated $10$ different sets of $200$ SLDs each. Table \ref{SLD_Results} summarizes the results obtained for the different sets of SLDs. Figure \ref{Fig:RegDistributionSLD} shows the \emph{median} distribution of the deployed regenerators over the network nodes. We can notice that for small values of $\pi$ ($\pi\in\{0.1,0.2,0.3\}$), the $M:N$ and $1+1$ protection schemes achieve the same results and the nodes \node{5} and \node{10} are the only regeneration sites. For large values of $\pi$ ($\pi\in\{0.4,0.5,0.75\}$), nodes \node{4}, \node{6}, \node{9} and \node{10} host more than $70\%$ of the deployed regenerators. Moreover, for the latter values, the reduction in the number of deployed regenerators varies between $23\%$ and $30\%$ when comparing the $M:N$ and $1+1$ protection schemes.

\begin{table}
\centering{\footnotesize\setlength{\extrarowheight}{1pt}
\caption{Results for various loads of permanent PLDs.}\label{PLD_Results}\vspace{-6pt}
\begin{tabular}{c@{~}c|c@{~~}c|c@{~~}c|c@{~~}c}
 & $D$ & $\boldsymbol{\bar{\mathfrak{a}}}$ & $\boldsymbol{\ddot{\mathfrak{a}}}$ & $\boldsymbol{\bar{\phi}}$ & $\boldsymbol{\ddot{\phi}}$ & $\boldsymbol{\bar{\mathfrak{R}}}$ & $\boldsymbol{\ddot{\mathfrak{R}}}$\\
\hline
\multirow{3}{*}{\rotatebox[origin=c]{90}{\parbox[c]{1cm}{{\scriptsize\centering $\;M:N\;$ protection}}}} &
  100 & 100\% & 0\% & 2 & 0 & 42 & 4 \\
& 200 & 100\% & 0\% & 4.33 & 0.58 & 48.67 & 6.81 \\
& 300 & 87.56\% & 1.07\% & 9.33 & 1.15 & 87.33 & 13.05 \\ \hline
\multirow{3}{*}{\rotatebox[origin=c]{90}{\parbox[c]{1cm}{{\scriptsize\centering  $\;1+1\;$ protection}}}} &
  100 & 100\% & 0\% & 1 & 0 & 21 & 2 \\
& 200 & 100\% & 0\% & 2.67 & 0.58 & 31.33 & 3.51 \\
& 300 & 88.33\% & 1.67\% & 4 & 0 & 59.67 & 8.08 \\ \hline
\end{tabular}}\vspace{6pt}
\centering{\footnotesize\setlength{\extrarowheight}{1pt}
\caption{Results for for various sets of dynamic SLDs.}\label{SLD_Results}\vspace{-6pt}
\begin{tabular}{c@{~}c|c@{~~}c|c@{~~}c|c@{~~}c}
 & $\pi$ & $\boldsymbol{\bar{\mathfrak{a}}}$ & $\boldsymbol{\ddot{\mathfrak{a}}}$ & $\boldsymbol{\bar{\phi}}$ & $\boldsymbol{\ddot{\phi}}$ & $\boldsymbol{\bar{\mathfrak{R}}}$ & $\boldsymbol{\ddot{\mathfrak{R}}}$\\
\hline
\multirow{6}{*}{\rotatebox[origin=c]{90}{\parbox[c]{1cm}{{\scriptsize\centering $\;M:N\;$ protection}}}} &
 0.75 & 100\% & 0\% & 5.67 & 0.58 & 68 & 7 \\
& 0.5 & 100\% & 0\% & 5.33 & 0.58 & 65.33 & 16.44 \\
& 0.4 & 100\% & 0\% & 3 & 0 & 43.33 & 10.21 \\
& 0.3 & 100\% & 0\% & 2 & 0 & 40.67 & 6.43 \\
& 0.2 & 100\% & 0\% & 2 & 0 & 28.67 & 6.43 \\
& 0.1 & 100\% & 0\% & 2 & 0 & 18.67 & 4.16 \\ \hline
\multirow{6}{*}{\rotatebox[origin=c]{90}{\parbox[c]{1cm}{{\scriptsize\centering  $\;1+1\;$ protection}}}} &
 0.75 & 100\% & 0\% & 3 & 0 & 48.33 & 8.39 \\
& 0.5 & 100\% & 0\% & 3 & 0 & 42.67 & 5.51 \\
& 0.4 & 100\% & 0\% & 1.67 & 0.58 & 28.67 & 6.43 \\
& 0.3 & 100\% & 0\% & 1 & 0 & 20.33 & 3.21 \\
& 0.2 & 100\% & 0\% & 1 & 0 & 14.33 & 3.21 \\
& 0.1 & 100\% & 0\% & 1 & 0 & 9.33 & 2.08 \\ \hline
\end{tabular}}
\end{table}

Finally, it should be noted that nodes \node{3}, \node{7}, \node{8} and \node{12} were seldom selected as regeneration sites.

\section{Conclusion}\label{Conclusions}

Reducing the number of regenerators and regeneration sites is highly motivated by the reduction in power consumption and maintenance cost. However, excessively concentrating the regenerators into a small number of nodes exposes the network to a high risk of data losses in the hazardous event of a regenerator pool failure. Thus, it is essential to keep in mind the network survivability concern while dimensioning the network. In this paper, we propose an exact approach based on a mathematical formulation that implements an $M:N$ shared regenerator pool protection scheme. For slightly loaded network, the proposed approach achieves comparable results to the commonly deployed $1+1$ protection scheme. However, as the network load increases, the gain obtained by the $M:N$ protection scheme becomes more perceptible as the reduction in the number of deployed regenerators may exceed 25\%.



\bibliographystyle{IEEEtran}
\bibliography{IEEEfull,Bibliography}

\begin{thebibliography}{10}
\providecommand{\url}[1]{#1}
\csname url@samestyle\endcsname
\providecommand{\newblock}{\relax}
\providecommand{\bibinfo}[2]{#2}
\providecommand{\BIBentrySTDinterwordspacing}{\spaceskip=0pt\relax}
\providecommand{\BIBentryALTinterwordstretchfactor}{4}
\providecommand{\BIBentryALTinterwordspacing}{\spaceskip=\fontdimen2\font plus
\BIBentryALTinterwordstretchfactor\fontdimen3\font minus
  \fontdimen4\font\relax}
\providecommand{\BIBforeignlanguage}[2]{{%
\expandafter\ifx\csname l@#1\endcsname\relax
\typeout{** WARNING: IEEEtran.bst: No hyphenation pattern has been}%
\typeout{** loaded for the language `#1'. Using the pattern for}%
\typeout{** the default language instead.}%
\else
\language=\csname l@#1\endcsname
\fi
#2}}
\providecommand{\BIBdecl}{\relax}
\BIBdecl

\bibitem{SCHMIDT:2008}
T.~Schmidt, C.~Malouin, R.~Saunders, J.~Hong, and R.~Marcoccia, ``Mitigating
  channel impairments in high capacity serial {40 G} and {100 G DWDM}
  transmission systems,'' in \emph{Digest of the IEEE/LEOS Summer Topical
  Meetings}, 2008, pp. 141--142.

\bibitem{SHEN:2002}
G.~Shen, W.~Grover, T.~Cheng, and S.~Bose, ``Sparse placement of electronic
  switching nodes for low-blocking in translucent optical networks,'' \emph{OSA
  JON}, vol.~1, no.~12, pp. 424--441, Dec. 2002.

\bibitem{YOUSSEF:2011}
M.~Youssef, S.~Al~Zahr, and M.~Gagnaire, ``Translucent network design from a
  {CapEx/OpEx} perspective,'' \emph{Springer PNC}, vol.~22, no.~1, pp. 85--97,
  Aug. 2011.

\bibitem{YANG:2005}
X.~Yang and B.~Ramamurthy, ``Sparse regeneration in translucent
  wavelength-routed optical networks: architecture, network design and
  wavelength routing,'' \emph{Springer PNC}, vol.~10, no.~1, pp. 39--50, Jul.
  2005.

\bibitem{ALZAHR_ConTEL:2007}
S.~Al~Zahr, N.~Puech, and M.~Gagnaire, ``Gain equalization versus electrical
  regeneration tradeoffs in hybrid {WDM} networks,'' in \emph{Proc. of IEEE
  ConTel}, 2007, pp. 33c--44c.

\bibitem{PACHNICKE:2008}
S.~Pachnicke, T.~Paschenda, and P.~Krummrich, ``Assessment of a
  constraint-based routing algorithm for translucent {10 Gbits/s DWDM} networks
  considering fiber nonlinearities,'' \emph{OSA JON}, vol.~7, no.~4, pp.
  365--377, Apr. 2008.

\bibitem{PAN:2008}
Z.~Pan, B.~Chatelain, D.~Plant, F.~Gagnon, C.~Tremblay, and E.~Bernier, ``Tabu
  search optimization in translucent network regenerator allocation,'' in
  \emph{Proc. of IEEE BROADNETS}, 2008, pp. 627--631.

\bibitem{ZHANG:2009}
W.~Zhang, J.~Tang, K.~Nygard, and C.~Wang, ``Repare: Regenerator placement and
  routing establishment in translucent networks,'' in \emph{Proc. of IEEE
  GLOBECOM}, 2009, pp. 1--7.

\bibitem{MANOU-JLT:2009}
K.~Manousakis, K.~Christodoulopoulos, E.~Kamitsas, I.~Tomkos, and
  E.~Varvarigos, ``Offline impairment-aware routing and wavelength assignment
  algorithms in translucent {WDM} optical networks,'' \emph{IEEE/OSA JLT},
  vol.~27, no.~12, pp. 1866--1877, June 2009.

\bibitem{DOUMITH_ICC:2011}
E.~A. Doumith, S.~Al~Zahr, and M.~Gagnaire, ``Mutual impact of traffic
  correlation and regenerator concentration in translucent {WDM} networks,'' in
  \emph{Proc. of IEEE ICC}, 2011, pp. 1--6.

\bibitem{ALZAHR_ICT:2011}
S.~Al~Zahr, E.~A. Doumith, and M.~Gagnaire, ``An exact approach for translucent
  {WDM} network design considering scheduled lightpath demands,'' in
  \emph{Proc. of IEEE ICT}, 2011, pp. 450--457.

\bibitem{GAGNAIRE_ONDM:2011}
M.~Gagnaire, E.~A. Doumith, and S.~Al~Zahr, ``A novel exact approach for
  translucent {WDM} network design under traffic uncertainty,'' in \emph{Proc.
  of IEEE ONDM}, 2011, pp. 1--6.

\bibitem{AZODOLMOLY:2009}
S.~Azodolmolky, M.~Klinkowski, E.~Marin, D.~Careglio, J.~Sol\'e-Pareta, and
  I.~Tomkos, ``A survey on physical layer impairments aware routing and
  wavelength assignment algorithms in optical networks,'' \emph{Elsevier
  Comput. Netw.}, vol.~53, no.~7, pp. 926--944, May 2009.

\bibitem{MOREA:2008}
A.~Morea, N.~Brogard, F.~Leplingard, J.-C. Antona, T.~Zami, B.~Lavigne, and
  D.~Bayart, ``{QoT} function and a* routing: an optimized combination for
  connection search in translucent networks,'' \emph{OSA JON}, vol.~7, no.~1,
  pp. 42--61, Jan. 2008.

\end{thebibliography}

\end{document}